# The impossibility of one-way superluminal signaling in the framework of modern Physics

Moses Fayngold
Department of Physics, New Jersey Institute of Technology, Newark, NJ 07102

This is a response to the note criticizing my paper "A New Paradox in Superluminal Signaling." The criticism is based, among other things, on arbitrarily postulated statement about invariance of direction of superluminal information transfer. The analysis presented in my response shows the absence of any grounds for such statement and confirms my original proof of the impossibility of one-way superluminal signaling. The analysis also unveils many other inconsistencies in the discussed note and shows that all its criticism is unsubstantiated.

**Introduction**

I am pleased that my work [1] proving the impossibility of one-way superluminal signaling (SS) has been criticized [2]. The critique prompted me to review and take a closer look at my arguments. But a thorough revision presented below has not found any serious flaws in them. On the other hand, there are grave inconsistencies in [2], and discussing them will be the main subject of my response.

To begin with, the basic result obtained in [1] is totally consistent with the Sommerfeld-Brillouin theorem in the wave theory [3, 4]. In contrast, the conclusion in [2] that one-way SS remains a possibility clashes with this theorem.

The author of [2] also decrees "...*the principle of the invariance of the information flow direction*". No such principle exists. Some authors do argue for invariance of SS's direction (see, e.g., [5]), but the arguments do not necessarily make a principle. In the given case, the controversy originates from confusing two totally different characteristics of SS: the *amount* /*contents* of prepared information on the one hand, and its *propagation direction*, on the other. The first feature is associated with the pre-conditions in the signal emitter and receptor. The second feature is about signal *propagation* and satisfies the crystal-clear definition: *the direction of information transfer is the direction of motion of information carrier*. According to Special Relativity (SR), the emission /reception are interchangeable concurrently with reversal of information flow.

But since these simple rules have, for some mysterious reasons, escaped a big part of the Physics community, it warrants a thorough discussion. Accordingly, my comments consist of two parts. The first part (Sec.1) describes the preconditions for signaling and then analyzes the signal transfer between two agents A and B. This will confirm all results of [1]. The second part (Sec.2) presents the detailed analysis of work [2] and shows its inner inconsistencies making its claims unsubstantiated.

## 1. Preparation and transfer of superluminal information

Following [1], we identify the SS-carriers with hypothetical tachyons [6-11]. We consider a tachyon interaction with two objects A and B, denoting an *object* by a straight capital symbol, and an *event* in its history – by the corresponding *italics* capital.

The A and B are assumed to recede from each other with relative velocity $V$, so we can associate with them two respective inertial reference frames (RF) – one co-moving with A (frame K) and the other co-moving with B (frame K′). Since a tachyon's world-line is space-like, the time ordering of events $(A, B)$ in K′ is opposite to that in K when [10–15]

$$V \tilde{v} > c^2 \qquad (1)$$

Here $\tilde{v}$ is tachyon's velocity and the vectors $\mathbf{V}$, $\tilde{\mathbf{v}}$ are assumed to be collinear.

The fact that signal transfer from A to B in K is, under condition (1), observed as going from B to A in K′ can be expressed symbolically as:

$$A \xrightarrow[\tau,\text{ in K}]{} B \quad \text{and} \quad A \xleftarrow[\tau',\text{ in K}']{} B \qquad (2)$$

Both expressions in (2) describe *the same* one-way SS *with the same tachyon* $\tau$ but observed from two different RF and denoted as $\tau'$ in K′. They express mathematically the reinterpretation principle (RIP) for SS satisfying condition (1) [6-8].

In all such cases, cause and effect swap their roles: event $A$ is the cause of $B$ in K, but is its effect in K′. This saves the causal ordering postulate (COP) (cause precedes the effect) but the definition of cause and effect loses its Lorentz-invariance in SS.

Impossibility of SS had been proven for a two-ways cycle message-response. Under conditions (1) in each way, the response comes to the sender before the emission of the original signal, thus creating the Tolman paradox [12]. The paradox was precluded by banning "tachyon exchange". As Nick Herbert put it, "*Some physicists, noting that all time-travel paradoxes arise from returning to a location before you left it, decided to eliminate the paradox by refusing to issue round-trip tickets to tachyons*" [13].

But as for *one-way* SS, the major (albeit not unanimous) consensus allows it despite the result [3]. Apart from [3], there was, to our knowledge, no rigorous proof of the general impossibility of SS. Such a proof was presented in [1] and will be reviewed here. We will first describe some features of message creation (Sec.1.1) and then its transmitting between A and B (Sec.1.2). Then it will be shown in Sec. 1.3 that already one-way SS leads to contradictions, so SS must be forbidden in principle.

### 1.1 Some basic features of information creation

Partial acceptance of one-way SS within modern Physics is based on a widely spread misconception that the direction of information flow can be determined by its contents. In the known example [5], a message is the text of "Hamlet" sent by Shakespeare at A as a chain of tachyons to Bacon at B. Under condition (1), it would be recorded in K′ as emitted from B to A long before Shakespeare's birth. The corresponding conclusion was (cited in italics): "*…no amount of reinterpretation will make Bacon the author of Hamlet. It is Shakespeare, not Bacon, who exercises control over the content of the message. For* *any tachyon trajectory the time ordering of the end points is relative...But the direction of information transfer is necessarily a relativistic invariant*."

Now turn from Literature / Psychology to Physics. A physical characteristic of information is the number of emitted /received bits which is the same for A and B if the transmission is lossless. It has nothing to do with text's importance for some humans.

The complete communication imposes specific pre-conditions *at either end of the path*. In tachyon-tardyon dynamics, even the ground atomic states could spontaneously get excited by *emitting* a tachyon at the cost of the initial kinetic energy. Similarly, a moving excited atom could go to a lower energy level by absorbing a tachyon [1, 14, 15]. Accordingly, for the message to be received, the B atoms must be in appropriate excited states at appropriate moments of time. Only the appropriate tuning would allow successful transmission of a text between A and B. In no way the historical records of Shakespeare's authorship can prove the possibility of only $A \to B$ transfer of the respective information. The direction of information flow is determined exclusively by RIP. The Info created at A is sent to B according to the A-records. The same Info is preconditioned at B by creating lots of excited atoms; this allows its timely emission to A according to B-records, even if it appears to be emitted spontaneously there.

Turning back to conditions (2), consider the message's origin at either end in more details. Introduce observer Alice in K and Bob in $K'$. Suppose the objects A, B are the atoms with the ground and excited states $|0\rangle_A$, $|1\rangle_A$ and $|0\rangle_B$, $|1\rangle_B$, respectively.

The events at the *end points* of path *AB* may, as mentioned above, look strange when observed from the other end point. When Alice observes *both* events from her RF, tachyon emission by A in the optical-type transition $|1\rangle_A \to |0\rangle_A$ looks quite natural for her. Atom B *absorbs* this tachyon, but contrary to Alice's intuition, the respective transition will be $|1\rangle_B \to |0\rangle_B$ instead of $|0\rangle_B \to |1\rangle_B$. The energy released in this transition plus energy of the absorbed tachyon both add to kinetic energy of receding B. Since atom's "personal" history is frame-independent, the same transitions will be recorded in $K'$. But there, $|1\rangle_B \to |0\rangle_B$ is accompanied by tachyon *emission* and this looks natural to Bob.

By the same token, while $|1\rangle_A \to |0\rangle_A$ is accompanied by tachyon emission in K, Bob records tachyon absorption, with the same explanation as absorption at B seen by Alice. Two emissions – one from A as observed in K, and the other from B as observed in $K'$, are totally equivalent, and the same is true for absorptions. This reflects the symmetry in the initial conditions – each atom is initially at rest and properly tuned in respective RF.

Thus, with RIP accepted, both events *A* and *B* are causally equivalent. In K, event *A* initiates the transaction described by the left Eq. (2). In $K'$, this transaction (the right of Eq. (2)) is time-reversed and is initiated by *B*.

**1.2 The information flow in SS**

The problems with SS start already with a single tachyon. As shown in [16-18], tachyons are unreliable even for one-way signaling. But this falls short of a strict ban, so we continue the review of [1] giving *rigorous proof* of the impossibility of one-way SS.

Consider a chain of *n* bits, numbered as 1, 2,..., *j*,..., *n* and forming an extended message. Assume that each bit is a tachyon in one of the two possible states. The emission of each bit by A is characterized by the corresponding 4-vector $s_A^j \equiv (t_A^j, \mathbf{r}_A)$. Here $t_A^j$, $\mathbf{r}_A$ are, respectively, the moment of emission of *j*-th bit and its position at this moment in K. The $\mathbf{r}_A = const$ since A is stationary in K. Denote the corresponding absorptions at B as $s_B^j \equiv (t_B^j, \mathbf{r}_B^j)$. The same notations hold for the frame K′ with $t$, $\mathbf{r}$ primed, and the message is observed there as emitted by B.

We must distinguish between the time ordering of events $\mathbf{s}_A^j$, $\mathbf{s}_B^j$ for a *fixed* bit No. *j* and time ordering of consecutive bits in succession. The former is frame-dependent, whereas the latter is Lorentz-invariant. Suppose, A sends to B a superluminal message "*Nothing will come out of nothing*". The "emission-absorption" $\left(\mathbf{s}_A^j, \mathbf{s}_B^j\right)$ *of the same bit* is time-reversed in K′ under condition (1). On the other hand, the ordering of emissions (or absorptions) of two consecutive bits, say, $\left(\mathbf{s}_A^j, \mathbf{s}_A^{j+1}\right)$ or $\left(\mathbf{s}_B^j, \mathbf{s}_B^{j+1}\right)$, is Lorentz-invariant. In K′, the tachyons are departing from B, but their succession still reads "*Nothing will come out of nothing*". A superluminal message is robust even when departure-arrival ordering of each separate bit is time-reversible. Symbolically, using the above-introduced notations, we have

$$\left. \begin{array}{l} t_A^j < t_B^j \\ \Delta t_A > 0, \ \Delta t_B > 0 \end{array} \right\} \text{in K}, \quad \text{and} \quad \left. \begin{array}{l} t'^j_A > t'^j_B \\ \Delta t'_A > 0, \ \Delta t'_B > 0 \end{array} \right\} \text{in K}', \qquad (3)$$

where $\Delta t$ stands for the time interval of the whole message. As emphasized before, direction of the respective information flow is the direction of motion of its carriers, in the simplest case – one tachyon. Identifying the source by a special label as suggested in [2] will not change anything. According to RIP, the identifier will under condition (1) move together with the whole message from A to B in K and from B to A in K′. The relations (3) already invalidate all the arguments of [2].

The message's length (number of bits) and its *propagation* is presented graphically in Fig. 1. The red dotted line here depicts the first tachyon of the succession emitted by A and identifying A as the primary source, according to suggestion [2]. But it remains the first in the same succession when recorded in K′. One can as well declare it the evidence of priority of B. Each end acts as the emitter or receiver, depending on RF. A tachyon emission /absorption is *not* Lorentz-invariant.

This reversibility must also apply to the evolution of the tachyon itself during its travel between A and B. Its evolution observed in B would be time reverse of that observed in A [1]. This general rule cannot be changed by invoking entropy as a measure of time for a composite tachyon. The time direction is not determined by system's entropy [19]. A tachyonic spaceship with its inside dilapidating in K [20], would evolve from dilapidated to freshly painted in K′. And the respective records from both – K and K′ – would be equally legitimate. In all cases, they have certain symmetry with respect to each other in terms of both – laws of nature and the local initial conditions.

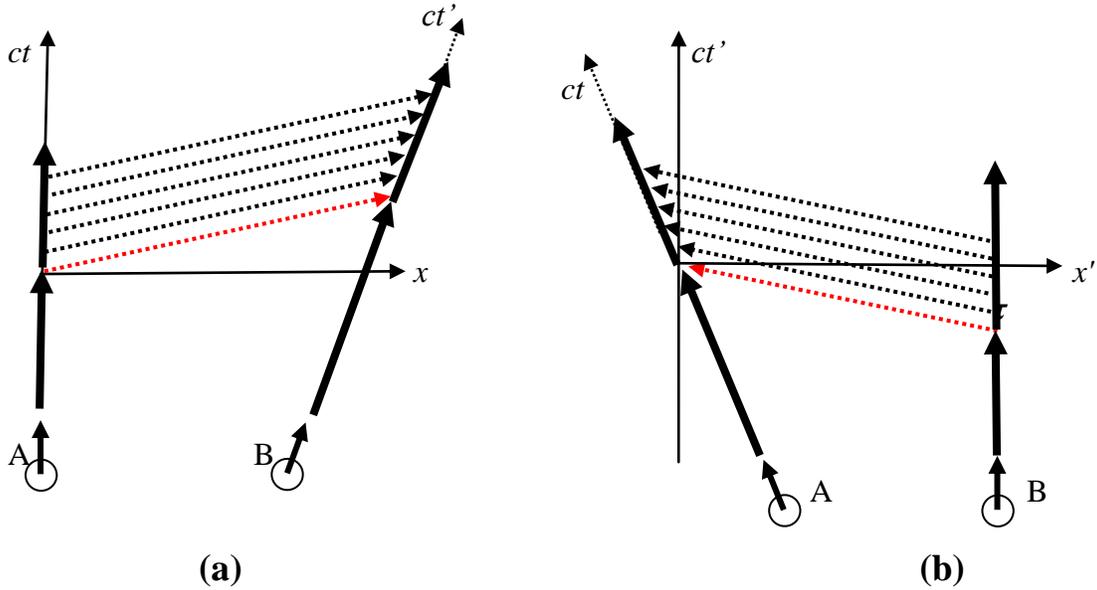

**Fig. 1**
Space-time diagrams of objects A and B connected by a structured tachyon signal (not to scale)
(a) – as observed in K;  (b) – as observed in K′.
The solid lines are the world-lines of the objects. The dotted lines are the world-lines of the messengers (tachyons). The recoils due to interaction with tachyons are neglected. In each RF only the temporal axis of the other frame is shown.

   To simplify the following discussion, we reduce, in contrast with suggestion [2], a succession shown in Fig.1 to a single tachyon. We can then consider a tachyon emission in either frame as a spontaneous event in that frame. In the absence of emission from A there are no emissions from B, and vice-versa.
   With all their weirdness, the discussed features of SS under described conditions are not self-contradictory, so they do not ban one-way communication. The strict ban is discussed in the next section.

### 1.3   Insertion of an intermediate absorber
   Here we review the rigorous proof of impossibility of one-way SS [1]. Its essence is the analysis of tachyon interaction with some other object in the path (AB). To this end, introduce a third party, Celia, whose world line C runs between A and B (Fig.2). On the verge of crossing the tachyon world line *AB* Celia puts an absorber C (an opaque plate) in the tachyon path. The resulting absorption (event *C*) happens when the tachyon is *indisputably on its way* between A and B. In K, the B-bound tachyon from A is absorbed by C ( $A \rightarrow C$, Fig. 2a), so the part *C*B does not materialize. Also, A and C are kicked away from each other. The pair of involved objects is (A, C), while B remains idle. Due to factual invariance of events, the same pair should be recorded by Bob, only in the reversed order ( $A \leftarrow C$, Fig. 2b). But from Bob's perspective, event *C* happens well *after B*, so it must block segment *C*A without affecting the preceding part *B*C.

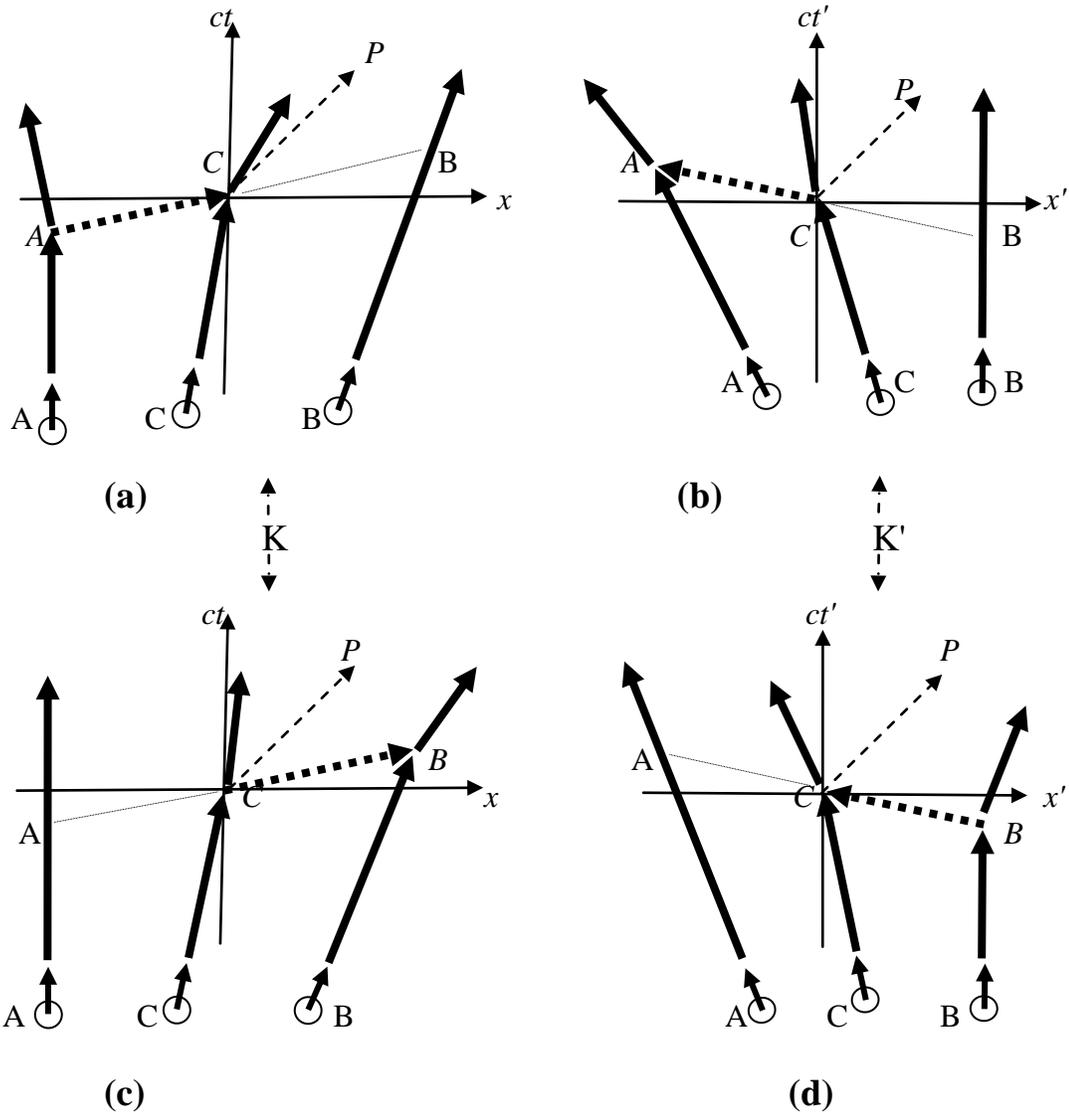

**Fig. 2** (From work [1])
Objects A and B connected by a single tachyon. The tachyon's interception by an intermediate absorber C produces paradoxical situation: neither of possible outcomes is compatible with known physical principles ($CP$ – the world line of a photon).
(a), (b) – Alice's expectations of the outcome (as seen in frames K and $K'$, respectively)
(c), (d) – Bob's expectations of the outcome (also as recorded in K and $K'$).
Their predictions are mutually exclusive.

To take a closer look, suppose that something distracts Celia right before her action. Then the tachyon just passes C, so Alice observes it traversing the whole path $A \to C \to B$ while Bob records $A \leftarrow C \leftarrow B$. Denoting the corresponding K-moments as $t_A$, $t_C$, and $t_B$, we have $t_A < t_C < t_B$. In $K'$, using primed notations, we have $t'_B < t'_C < t'_A$. The part $C \leftarrow B$ of the tachyon's trajectory is as real for Bob as is part $A \to C$ for Alice.

Therefore, when Celia does insert C at $t' = t'_C$, Alice claims it cancels *already pre-existing* part $C \leftarrow B$ of tachyon's history in $K'$, including the atomic transition $|1\rangle_B \to |0\rangle_B$ with tachyon emission at $t' = t'_B < t'_C$. Canceling *already evolving* process by a future event is an outrageous violation of causality. Thus, Alice's records contradict those of Bob.

But Bob's view is the same blind venue. According to Bob, event *C* must block the part $A \leftarrow C$ of the process, so Alice can observe only the prior stage $C \leftarrow B$ in the reversed order, as $C \to B$. The pair of objects interacting with tachyon and accordingly kicked away from each other is (B, C) (Fig.2c, d), whereas A remains idle. But that leads to the same clash with causality from the viewpoint of K.

Both scenarios contradict each other, and since they both describe the factual features of the same process rather than its numerical characteristics which might be relative, this is a logical contradiction. And in contrast with the twin paradox, which is easily resolved by noticing non-equivalence of the two used RF, there is no resolution here since both used RF are totally equivalent.

Thus, the described thought experiment may have four formally possible outcomes:
(a) Both observers record the pair of events (*A*, *C*) (albeit in the opposite ordering)
(b) Both record the pair (*B*, *C*), also in opposite ordering
(c) Alice records (*A*, *C*), whereas Bob records (*B*, *C*)
(d) Alice records (*C*, *B*), whereas Bob records (*C*, *A*) (this is a purely formal possibility)
The outcome (a) violates causality in $K'$ – event *C* cancels the *preceding* stage $C \leftarrow B$. Outcome (b) violates causality in K: event *C* cancels preceding process $A \to C$.

Each of the outcomes (c) and (d) violates the factual invariance, according to which all observers must record *the same* pair of events.

All four possibilities contradict basic principles. Thus, SS leads to paradoxes already in one-way communications. Paraphrasing Nick Herbert, if we adhere to basic principles, *we must refuse to issue even one-way tickets to SS-carriers*. And this would disqualify tachyons from carrying any messages.

*Conclusion to this part*: Even though superluminal motions as such do exist and are consistent with relativity [4, 21-25], none of them can be harnessed for SS. The obtained result imposes a universal ban on SS independent of the Sommerfeld-Brillouin theorem. And it is stronger than the Tolman paradox which appears only in the round-trip communications.

## 2. Analysis of work [2]

In this section, we scrutinize the critique of work [1] by V. Perepelitsa [2]. The direct quotations from [2] will be in quotation marks and in italics. But some of my statements that I want to emphasize can also be in italics. References in [2] to my work as [15] will be changed to [1] to avoid any confusion.

Let us now consider the basic arguments in [2].

I) "*Indeed, if one follows strictly the orthodox principle of causality, the introduction of the plug, viewed from the frame A, removes the part of the tachyon world line from C to B. On the other hand, this action, viewed from the frame of the observer B... should* *remove the part from C to A. Such an ambiguity is considered in [1] as a logical paradox which can be used to ban faster-than-light particles and signals.*"

This quotation uses ambiguous terminology. The adjective "*orthodox*" when applied to a *principle* means "outdated". But as of today, causality principle remains the bedrock of physics. Yes, I do follow this conventional (which is anything but "*orthodox*") principle, with unavoidable corollary of impossibility of SS. Naming the above-described contradictions just "*an ambiguity*" amounts to ignoring the basics of SR and deliberately dismissing simple logics.

II) "*What is wrong in this construction? The answer is: it violates the principle of the invariance of the information flow direction. It is easy to prove that this principle holds in any process of the information transfer, whatever could be the time order of the sending and receiving of the information (as we have seen, this order can be reversed in the case of the tachyon exchange).*"

This is a mix of the wrong statements. First, as shown in Sec.1, there is no "*...principle of the invariance of the information flow direction*". Second, "*It is easy to prove...*" is not confirmed by any known and accepted proof. Third, the whole statement confuses the information flow *between* two agents with a process of its preceding creation in the respective device. Reversal of time ordering of the events (*A*, *B*) in SS is the reversal of information flow between them. This is especially obvious for each separate tachyon, with no artificial headache about totally irrelevant notion of "authorship."

Consider now the "arguments" claimed to be a proof of the information flow invariance:

III) "*For the proof we may turn our consideration from a single-tachyon signal to the signal containing much information (e.g. transmitted by a modulated tachyon beam). Then two straightforward arguments can be used in order to prove the invariance of the information flow direction. First, we note that any information message presents a sequence of symbols (e.g. letters and numbers) separated by time-like intervals. Therefore the time ordering of the symbols in this sequence is invariant, i.e. it does not change even when the message is sent with the faster-than-light (spacelike) carriers, which can go backward in time. Second, each information message can bear an identifier of its sender, so the source of the message can be determined without doubt in any reference frame, whatever could be the time order of the message sending and receiving. The application of the principle of the invariance of the information flow direction results in doubtless identification of a cause and its effect in any causally related sequence of events.*"

This is *anything but* proof. Actually, each statement here when taken separately just mirrors my analysis in [1] and in Sec.1.2 of this work, but the drawn conclusions are the opposite, in an obvious contradiction with logics.
First, a single tachyon is a simplest case to analyze. A bunch is a more complex system than each of its ingredients. Adding equally moving tachyons to the first one could at best only increase the *amount* of transferred information, but not affect its propagation direction. Turning from a single tachyon to a tachyon beam would only complicate rather than simplify the aimed proof. The only "gain" in choosing a succession of tachyons instead of a single one is an opportunity to invoke an unphysical notion of the "authorship" and to wrongly use it as a "proof" of invariance of SS direction. Nothing can be father from a proof than such an argument.

Second, bearing an identifier of the information sender does not in any way affect the *direction* of information flow. It only creates an illusion of one sender's priority, but an illusion is not a mathematical proof. *The same identifier* is emitted from B to A according to Bob's records, see Fig.1.

Ask any reader accepting RIP a simple question: Suppose you reside near B and observe each separate tachyon carrying one bit from B to A in total accord with RIP. Will you nevertheless assert that the string of such bits carries info from A to B? In other words, each single tachyon transfers some information from B to A, but the string of them reverses the information flow. This is precisely the assertion of [2] and it invites the questions: How many additional tachyons will one need to reverse this flow? And what is the mechanism of such reverse? No surprise that applying this kind of "*proof*" to situation considered in Sec.1 and in [1], the author of [2] obtains 4 nonsensical statements. Here is the corresponding quotation:

IV) "*Applying it to the consideration of a concrete one-way signaling described above, we arrive to four different situations, all of them being logically self-consistent*:
 a) If the observer A issues a superluminal signal and it is blocked by the observer C, i.e. at the position of C, all the observers record the pair of the events (A, C).
 b) If the observer B issues a superluminal signal and it is blocked at the position of C, all the observers record the pair (B, C).
 c) If the observer A issues a superluminal signal and it is not blocked by the observer C, i.e. it goes to B and is absorbed there, all the observers record the pair of the events (A, B).
 d) If the observer B issues a superluminal signal and it is not blocked at the position of C, all the observers record the pair (B, A)."

All four situations *a) – d)* are proclaimed by their author as "*logically self-consistent*." Actually, they are quite the opposite. Using Pauli's famous expression, they are *Not Even Wrong*. They are just *declared* by the author without any attempts to justify them.

Let us start with the basic result of reversibility of SS between two mutually receding objects under condition (1). The succession ($A \rightarrow B$) as recorded in frame A is ($A \leftarrow B$) in frame B. In view of the *equivalence of all inertial RF*, both records are equally legal. Applying this to *a) – d)* immediately shows the following.

The situation *a)* is nonsensical because the condition "*If*" cannot be entirely met in frame B. In that frame, such "*If*" applies *only* to blocking the signal at C but not to its emitting at A because the B-observer records in this situation the pair of events (*B, C*). So Bob physically cannot join the camp of "...*all the observers*". The similar argument applies to *b)*. *If* a superluminal signal is blocked at C, the observer A records the pair (*A, C*) *and* observer B records the pair (*B, C*). "...*All the observers*" cannot record *the same* pair in the given case.

In both statements *c)* , *d)*, the author contradicts himself. According to RIP, both observers do record in this case the same pair of events, but in the opposite ordering.

The correct statements are: If a superluminal signal is *not* blocked by C, it is recorded as pair (*A, B*) in frame A and as pair (*B, A*) in frame B. This is self-consistent, so logical contradictions banning one-way SS emerge only at blocking the signal on its way between the edges of the corresponding space-like interval.

V) "*Turning to the concluding section of [1] we can reformulate several statements related to the case when the observer C blocks the tachyon signal, not allowing a passage of tachyons through his plug, to make these statements correct (the corrections to the original items of [1] are* underlined here*):*
 (a) <u>*If the observer A issues a superluminal signal*</u> *both observers record the pair of the events (A, C) (albeit in the opposite ordering)*
 (b) <u>*If the observer B issues a superluminal signal*</u> *both record the pair (B, C), also in the opposite ordering.*

As in the previous quotation, the condition "*If*" here is *Not Even Wrong*. The condition *(a)* is not and cannot be met in frame B when SS is intercepted at C. *Both* observers do *not* and *cannot* record *the same* pair (*A, C*) because the event *A* is *precluded* in K′ by signal interception. For the same reason the "*correction*" in *(b)* is also invalid.

VI) " *The outcomes* (c) *and* (d) *considered in the concluding section of [1] should be discarded as mutually controversial.*"

Then by the same logics the author must discard the Tolman paradox because the response from B prevents the initial signaling from A, and these two events are "...*mutually controversial.*" According to [2], any argument invalidating some wrong assumption can be discarded as controversial precisely because of invalidating the assumption. This is not how logic works.

### *Conclusion*
The review of work [1] confirms all its basic arguments including the proof of impossibility of the one-way SS. On the other hand, the analysis of the arguments of work [2] shows that the respective criticism is totally unsubstantiated.
### *Acknowledgements*
I am grateful to Art Hobson and Nick Herbert for valuable comments and to Anwar Shiekh for thorough discussions.